\documentstyle[11pt,newpasp,twoside]{article}
\def\edcomment#1{\iffalse\marginpar{\raggedright\sl#1\/}\else\relax\fi}
\reversemarginpar

\begin{document}
\title{Prospects for future far-infrared/submillimeter
studies of the high-redshift Universe}
\author{Andrew W. Blain}
\affil{Institute of Astronomy, Madingley Road, Cambridge, CB3 0HA, UK} 
\begin{abstract}
Observations made using {\it COBE}, SCUBA, {\it ISO} and MAMBO have provided a 
reasonable working knowledge of both the intensity of the submm and far-infrared
background radiation and the source counts of luminous high-redshift dusty 
galaxies. However, because there are uncertainties in the background intensity
determinations, the samples of detected galaxies are small, and most importantly, 
their redshift distributions are very incomplete, details of the evolution of 
dusty galaxies remain unresolved. The next steps forward in the field will be the 
launches of {\it SIRTF} and {\it ASTRO-F}, the commissioning of SOFIA and new, more 
capable ground-based mm/submm-wave cameras -- BOLOCAM, SHARC-II and 
SCUBA-II -- the use of ultra-long duration balloon experiments, such as BLAST, the 
construction of ALMA and the arrival of {\it FIRST}, and ultimately the advent of 
space-borne far-infrared interferometers, such as {\it SPECS}. There are also 
exciting prospects for direct mm/submm-wave CO-line redshift surveys using 
wide-band spectrographs. Using these new facilities, the number of high-redshift 
dusty galaxies known will be increased dramatically. Spectroscopy using {\it SIRTF}, 
SOFIA and {\it FIRST} will probe the astrophysical processes within these sources 
in detail, hopefully addressing the open question of the fraction of the counts and 
background radiation that is generated by the formation of high-mass stars and by 
active galactic nuclei (AGNs). The spatial and spectral structure of 
distant dusty galaxies will finally be resolved in detail using 
ALMA and {\it SPECS}. 
\end{abstract}

\section{Introduction}

The determination of the intensity of extragalactic background radiation in the submm 
and far-infrared wavebands has been a very important development in
observational cosmology (Puget et al.\ 1996; Hauser et al.\ 1998; Schlegel, 
Finkbeiner \& Davis 1998; Finkbeiner, Davis, \& Schlegel 2000). Over the same period, 
the first observations that were sufficiently deep to detect high-redshift galaxies in the 
mm, submm and far-infrared wavebands were made using the 1.25-mm MAMBO 
bolometer camera at the IRAM 30-m telescope (Bertoldi et~al.\ 2000), the 450/850-$\mu$m 
SCUBA camera at the JCMT (Smail, Ivison, \& Blain 1997; for a current 
summary see 
Blain et al.\ 2000a) and the PHOT 
instrument aboard {\it ISO} (Kawara et al.\ 1998; Puget et al.\ 1999; Juvela, Mattila, \& 
Lemke 2000). Most of the results for both counts and background have been 
summarized elsewhere, see for example Figs\,9 and 12 in Blain et al.\ (1999b). 
Progress has been rapid: not only has the background radiation spectrum been 
determined for the first time, but most of the 850-$\mu$m background radiation, 
and about 30\% and 10\% of the background at wavelengths of 450 and 175\,$\mu$m 
respectively has been accounted for as coming from individual detected galaxies. 

This progress has been possible only because of tremendous improvements in the 
sensitivity and field of view of instruments in these wavebands. Many 
instrumental references can be found in Blain (1999a): to save space in this brief 
article they are generally not repeated here. The important pieces of missing 
information and the prospects for future developments of observations in the mm, 
submm and far-infrared wavebands are discussed below.

\section{Key missing information -- redshifts  of submm-selected galaxies} 

The mm--far-infrared background radiation spectrum is generated by thermal 
emission from interstellar dust in all the galaxies that lie along the line of sight, right 
back to the epochs when the first dust was formed during the early process of galaxy 
formation. It potentially includes emission from redshifts beyond 
recombination, as a neutral Universe is transparent to long-wavelength radiation. 
The intensity and spectral shape of the background radiation provides 
information about the volume emissivity, intrinsic spectral energy distribution 
and redshift of dusty galaxies; however, this information is not unambiguous. 
The counts of galaxies that make up the brighter fraction of the population 
contributing to the background can be used to refine and confirm the results of 
analyses of the background spectrum, but unfortunately neither the values of 
these quantities nor different scenarios of galaxy evolution can be distinguished 
strongly. Indeed, from the background and count data alone, the median redshift of 
galaxies detected in deep SCUBA images at 850-$\mu$m could plausibly lie between 
1 and 5. In order to discriminate between different well fitting models, measurements 
of redshift distributions are necessary; determinations of accurate counts 
at other mm/submm wavebands are useful, but less important. For more details 
see Blain et al.\ (1999a,b)

Unfortunately, the redshift distribution of SCUBA galaxies is hard to determine for 
two reasons. First, it is difficult to identify counterparts for submm-selected galaxies 
in other wavebands because their positions are relatively uncertain. The JCMT 
beam is 15\,arcsec wide, and so the centroids of even the most significant 
detections are uncertain to within several arcseconds. Secondly, counterparts 
are reasonably expected to be extremely faint, perhaps so faint that optical 
spectroscopy will remain challenging until {\it NGST} is available (Smail et al.\ 2000).

Deep radio observations are very useful for finding the positions of counterparts 
(Ivison et al.\ 1998; 2000). The flux density ratio between the submm and radio 
wavebands can also be used to provide an indication of redshift (Carilli \& Yun 1999, 
2000; Smail et al.\ 2000; Barger, Cowie, \& Richards 2000). Note however that the result 
is offset to lower values by the presence of any radio emission from a buried AGN, and 
depends on the dust temperature in the galaxy (Blain 1999b), with hotter galaxies at higher 
redshifts being difficult to distinguish from cooler ones closer by. Observations of
high-resolution dust emission using mm-wave interferometer arrays can also be used 
to obtain better positions (Downes et al.\ 1999; Frayer et al.\ 2000; Gear et al.\ 2000); 
however, accurate relative optical--mm astrometry may be difficult in the 
absence of deep radio data because of the small fields of both 
the optical and mm-array images.

To confirm an identification beyond reasonable doubt, it is necessary to obtain an 
optical/near-infrared redshift for the candidate and then detect CO line emission 
at the same redshift and position. However, this is a slow process and an accurate 
redshift is essential. So far only three submm-selected galaxies have been observed
in this way (Frayer 1998, 1999; Kneib et al.\ in prep), less than 10\% of those described in 
the literature. Future wide-band mm and submm spectrographs will improve this situation. 


\section{Future progress} 

At present, the SCUBA camera detects sources at the rate of about one per 8-hour 
shift at the JCMT. Within the next few years, new large-format mm--submm 
cameras on ground-based telescopes -- including BOLOCAM for the CSO/50-m LMT, 
SHARC-II (Dowell, Moseley,  \& Phillips 2000) for the CSO and SCUBA-II for the JCMT -- 
will offer more rapid detection rates and access new wavebands. At 
shorter wavelengths, the MIPS instrument onboard {\it SIRTF}, the HAWC 
camera aboard SOFIA and balloon-borne instruments such as BLAST (Devlin et al.\ 2000) 
will provide similar improvements over existing surveys. 
The simple increase in the detection rate of galaxies will assist follow-up studies 
by providing larger samples of objects to sift. In addition, the higher spatial resolution 
of images -- from SCUBA-II at 450\,$\mu$m, SHARC-II at 350\,$\mu$m, and BOLOCAM
at 1.1\,mm on the LMT -- will make the identification process easier. Wider 
survey fields and greater numbers of detections will allow large-scale statistical 
studies of the distribution and clustering of high-redshift dusty galaxies to be carried 
out (Haiman \& Knox 2000).

There are also excellent prospects for more rapid detection of redshifted CO emission 
from candidate galaxies, mainly by increasing the bandwidth of instruments from the 
current maximum of 4\,GHz to several 10's of GHz. This will be sufficient to yield a 
realistic chance of detecting simultaneously adjacent CO lines -- separated by 
$115/(1+z)$\,GHz in the observer's frame for a galaxy at redshift $z$ -- from a 
submm-selected galaxy, and so determine a redshift without requiring 
optical observations (see Blain et al.\ 2000b; Blain 2000). 

In the longer term, the ground-based ALMA interferometer array and space-borne 
mid- and far-IR interferometers such as {\it SPECS} will provide wide spectral coverage, 
sub-arcsec spatial resolution and great sensitivity. Using these instruments the detailed 
astrophysics within distant dusty galaxies will be resolved in detail. However, 
interferometers have small fields of view. Wide-field surveys made using the {\it FIRST} 
and {\it Planck Surveyor} space missions will be required provide extensive lists of 
targets, in addition to the legacy of observations left by surveys made using {\it SIRTF} 
and {\it ASTRO-F} (Takeuchi et al.\ 1999). 

\section{Summary} 

A wide range of new instruments for studying the evolution of galaxies in 
the mm, submm and far-infrared are proposed and under construction. Observations 
using these facilities will enhance significantly the size and spatial resolution of 
catalogs of galaxies selected in these wavebands, and the amount of spectroscopic 
information available. The goal of understanding the history of dust-enshrouded 
galaxy formation and evolution in detail by determining the redshift distributions of 
galaxies selected in these wavebands will be realized. 

\acknowledgments{
I thank the Raymond \& Beverly Sackler Foundation, and ESO for support at the 
IAU meeting.}


\begin{references} 
\reference Barger, A. J., Cowie L. L., \& Richards, E. A., 2000, AJ, 119, 2092
\reference Bertoldi, F., et al 2000, A\&A, 360, 92 
\reference Blain, A. W. 1999a, ASP Conf. Ser. Vol. 191, 255 (astro-ph/9906141)
\reference Blain, A. W. 1999b, MNRAS, 309, 955
\reference Blain, A. W., Smail I., Ivison R. J., \& Kneib J.-P. 1999a, MNRAS, 302, 632
\reference Blain, A. W., et al. 1999b, MNRAS, 309, 715
\reference Blain, A. W. 2000, ASP Conf. Ser., in press (astro-ph/9911449)
\reference Blain, A. W., et al. 2000a, ASP Conf. Ser. Vol. 193, 246 (astro-ph/9908111) 
\reference Blain, A. W., Frayer, D. T., Bock, J., \& Scoville, N. 2000b, MNRAS, 313, 559 
\reference Carilli, C. L., \& Yun, M. S. 1999, ApJ, 513, L13
\reference Carilli, C. L., \& Yun, M. S. 2000, ApJ, 530, 618
\reference Devlin, M., et al. 2000, Proc. UMass conference, in press
\reference Dowell, C. D., Moseley, S. H., \& Phillips, T. G. 2000, ASP
Conf. Ser., in press 
\reference Downes, D., et al. 1999, A\&A, 347, 809 
\reference Finkbeiner, D. P., Davis, M., et al. 
2000, ApJ, in press
(astro-ph/0004175) 
\reference Frayer, D. T., et al. 1998, ApJ, 506, L7
\reference Frayer, D. T., et al. 1999, ApJ, 514, L13
\reference Frayer, D. T., et al. 2000, AJ, in press (astro-ph/0005239)
\reference Gear, W. K., et al. 2000, MNRAS, in press (astro-ph/0007054) 
\reference Haiman, Z., \& Knox, L. 2000, ApJ, 530, 124
\reference Hauser, M. G., et al. 1998, ApJ, 508, 106
\reference Ivison, R. J., et al. 1998, MNRAS, 298, 583
\reference Ivison, R. J., et al. 2000, MNRAS, 315, 209 
\reference Juvela, M., Mattila, K., \& Lemke, D. 2000, A\&A, in press (astro-ph/0005564)
\reference Kawara, K., et al. 1998, A\&A, 336, L9  
\reference Puget, J.-L., et al. 1996, A\&A, 308, L5
\reference Puget, J.-L., et al. 1999, A\&A, 345, 29
\reference Schlegel, D. J., Finkbeiner, D. P., \& Davis, M. 1998, ApJ, 500, 525
\reference Smail, I., Ivison, R. J., and Blain, A. W. 1997, ApJ, 490, L5
\reference Smail, I., Ivison, R. J., Owen, F. N., et al. 2000, ApJ, 528, 612 
\reference Smail, I., et al. 2000, Proc. UMass conference (astro-ph/0008237)
\reference Takeuchi, T. T., et al. 1999, PASP, 111, 288
\end{references}
\end{document}